\documentstyle[sprocl,epsfig,rotate]{article}
\newcommand{\nwc}{\newcommand}
\nwc{\cl}  {$\clubsuit$}

\nwc{\hyp} {\hyphenation} 
\nwc{\be}  {\begin{equation}}
\nwc{\ee}  {\end{equation}}
\nwc{\ba}  {\begin{array}}
\nwc{\ea}  {\end{array}}
\nwc{\bdm} {\begin{displaymath}}
\nwc{\edm} {\end{displaymath}}
\nwc{\bea} {\be\ba{rcl}}
\nwc{\eea} {\ea\ee}
\nwc{\ben} {\begin{eqnarray}}
\nwc{\een} {\end{eqnarray}}
\nwc{\bda} {\bdm\ba{lcl}}
\nwc{\eda} {\ea\edm}
\nwc{\bc}  {\begin{center}}
\nwc{\ec}  {\end{center}}
\nwc{\ds}  {\displaystyle}
\nwc{\bmat}{\left(\ba}
\nwc{\emat}{\ea\right)}
\nwc{\non} {\nonumber}
\nwc{\bib} {\bibitem}
\nwc{\lra} {\longrightarrow}
\nwc{\Llra}{\Longleftrightarrow}
\nwc{\ra}  {\rightarrow}
\nwc{\Ra}  {\Rightarrow}
\nwc{\lmt} {\longmapsto}
\nwc{\prl} {\partial}
\nwc{\iy}  {\infty}
\nwc{\ol}  {\overline}
\nwc{\hm}  {\hspace{3mm}}
\nwc{\lf}  {\left}
\nwc{\ri}  {\right}
\nwc{\lm}  {\limits}
\nwc{\lb}  {\lbrack}
\nwc{\rb}  {\rbrack}
\nwc{\ov}  {\over}
\nwc{\pri}  {\prime}
\nwc{\nnn} {\nonumber \vspace{.2cm} \\ }
\nwc{\Sc}  {{\cal S}}
\nwc{\Lc}  {{\cal L}}
\nwc{\Rc}  {{\cal R}}
\nwc{\Dc}  {{\cal D}}
\nwc{\Oc}  {{\cal O}}
\nwc{\Cc}  {{\cal C}}
\nwc{\Pc}  {{\cal P}}
\nwc{\Mc}  {{\cal M}}
\nwc{\Ec}  {{\cal E}}
\nwc{\Fc}  {{\cal F}}
\nwc{\Hc}  {{\cal H}}
\nwc{\Kc}  {{\cal K}}
\nwc{\Xc}  {{\cal X}}
\nwc{\Gc}  {{\cal G}}
\nwc{\Zc}  {{\cal Z}}
\nwc{\Nc}  {{\cal N}}
\nwc{\fca} {{\cal f}}
\nwc{\xc}  {{\cal x}}
\nwc{\Ac}  {{\cal A}}
\nwc{\Bc}  {{\cal B}}
\nwc{\Uc}  {{\cal U}}
\nwc{\Vc}  {{\cal V}}
\nwc{\Th} {\Theta}
\nwc{\th} {\theta}
\nwc{\vth} {\vartheta}
\nwc{\eps}{\epsilon}
\nwc{\si} {\sigma}
\nwc{\Gm} {\Gamma}
\nwc{\gm} {\gamma}
\nwc{\bt} {\beta}
\nwc{\La} {\Lambda}
\nwc{\la} {\lambda}
\nwc{\om} {\omega}
\nwc{\Om} {\Omega}
\nwc{\dt} {\delta}
\nwc{\Si} {\Sigma}
\nwc{\Dt} {\Delta}
\nwc{\al} {\alpha}
\nwc{\vp} {\varphi}
\nwc{\vph}{\varphi}
\nwc{\kp} {\kappa}
\def\tr{\mathop{\rm tr}}
\def\Tr{\mathop{\rm Tr}}

\def\VEV#1{\left\langle #1\right\rangle}

\nwc{\Id}  {{\bf 1}}
\nwc{\diag} {{\rm diag}}
\nwc{\inv}  {{\rm inv}}
\nwc{\mod}  {{\rm mod}}
\nwc{\hal} {\frac{1}{2}}
\nwc{\tpi}  {2\pi i}

\def\slash#1{#1\!\!\!/\!\,\,}

\def\plb#1{Phys.\ Lett.\ {\bf B#1}}

\def\prd#1{Phys.\ Rev.\ {\bf D#1}}

\def\NP#1{Nucl.\ Phys.~{\bf #1}}

\def\PR#1{Phys. Rev.~{\bf #1}}

\def\ZP#1{Z. Phys.~{\bf #1}}

\def\MeV {\,{\rm  MeV}}

\def \lta {\mathrel{\vcenter
     {\hbox{$<$}\nointerlineskip\hbox{$\sim$}}}}

\newsavebox{\nnin} \sbox{\nnin}{$\hspace{1mm}\in\kern -.8em /
                   \hspace{1mm}$}

\newcommand{\sub}{\subset}
\newsavebox{\nnsub} \sbox{\nnsub}{$\hspace{1mm}\sub\kern -.9em /
            \hspace{1mm}$}

\def\KK{{\rm I\kern -.2em  K}}
\def\NN{{\rm I\kern -.16em N}}
\def\RR{{\rm I\kern -.2em  R}}
\def\ZZ{Z \kern -.43em Z}
\def\QQ{{\rm \kern .25em
             \vrule height1.4ex depth-.12ex width.06em\kern-.31em Q}}
\def\CC{{\rm \kern .25em
             \vrule height1.4ex depth-.12ex width.06em\kern-.31em C}}
\def\ZZZ{Z\kern -0.31em Z}

\nwc{\olnu}  {\ol{\nu}}
\nwc{\olla}  {\ol{\la}}
\nwc{\olm}   {\ol{m}}
\nwc{\olmu}  {\ol{\mu}}
\nwc{\olh}   {\ol{h}}
\nwc{\olpsi} {\ol{\psi}}
\nwc{\olsi}  {\ol{\sigma}}
\nwc{\olgm}  {\ol{\gm}}
\nwc{\prlt}  {\frac{\prl}{\prl t}}
\nwc{\ttau}  {\tilde{\tau}}
\nwc{\trho}  {\tilde{\rho}}
\nwc{\tP}    {\tilde{P}}
\nwc{\tU}    {\tilde{U}}
\nwc{\teps}  {\tilde{\eps}}
\nwc{\tla}   {\tilde{\la}}
\nwc{\tit}    {\tilde{t}}
\nwc{\iddq}  {\int\frac{d^dq}{(2\pi)^d}}
\nwc{\prpr}  {\prime\prime}
\nwc{\rN}    {\left(\frac{\rho}{N}\right)}
\nwc{\rNt}    {\left(\frac{\rho}{N}\right)^{\frac{N-2}{2}}}
\nwc{\rnN}   {\left(\frac{\rho_0}{N}\right)}
\nwc{\rnNt}    {\left(\frac{\rho_0}{N}\right)^{\frac{N-2}{2}}}
\nwc{\rnNf}    {\left(\frac{\rho_0}{N}\right)^{\frac{N-4}{2}}}
\nwc{\rNs}    {\left(\frac{\rho_0}{N}\right)^{\frac{N-6}{2}}}
\nwc{\kNt}    {\left(\frac{\kappa}{N}\right)^{\frac{N-2}{2}}}
\nwc{\kNf}    {\left(\frac{\kappa}{N}\right)^{\frac{N-4}{2}}}
\nwc{\kNs}    {\left(\frac{\kappa}{N}\right)^{\frac{N-6}{2}}}

\begin{document}
\title{CHIRAL PHASE TRANSITION FROM NON--PERTURBATIVE FLOW 
EQUATIONS \footnote{Talk given at Eotvos Conference in Science: 
Strong and Electroweak Matter, Eger, Hungary, 21-25 May 1997.}} 
\author{J{\"U}RGEN BERGES}
\address{Institut f{\"ur} Theoretische Physik, Universit{\"a}t Heidelberg, 
Philosophenweg 16, D--69120 Heidelberg, Germany}     
\maketitle

\begin{picture}(5,2.5)
\put(240,120){HD--THEP--97--42}
\end{picture}

\abstracts{We employ non--perturbative flow equations 
to compute the equation of 
state for two flavor QCD within an effective quark meson model. 
Our treatment covers both the chiral perturbation
theory domain of validity and the domain of validity of universality
associated with critical phenomena. 
In the vicinity of $T_c$ and zero quark mass we obtain a
precision estimate of the universal critical equation of state
of the three dimensional $O(4)$ symmetric Heisenberg model.
For realistic quark masses the
pion correlation length near $T_c$
turns out to be smaller than its zero temperature value. }

QCD in a thermal equilibrium situation
at sufficiently high temperature differs in
important aspects from the corresponding zero temperature or
vacuum properties \cite{CP75}. A phase transition at some critical
temperature $T_c$ or a relatively sharp crossover may separate the
high and low temperature physics. Concentrating on the 
chiral aspects of QCD the transition is related to a qualitative
change in the chiral condensate. It was pointed
out \cite{PW84-1,RaWi93-1} that for sufficiently small up and 
down quark masses, $m_u$ and
$m_d$, and for a sufficiently large mass of the strange quark, $m_s$,
the chiral transition is expected to belong to the universality class
of the three dimensional $O(4)$ Heisenberg model. It
was suggested~\cite{RaWi93-1} that a large correlation length
may lead to a disoriented
chiral condensate~\cite{Ans88-1} with possible distinctive
signatures~\cite{RaWi93-1} in a relativistic heavy--ion collision.
The question how
small $m_u$ and $m_d$ would have to be in order to see a large
correlation length near $T_c$ and if this scenario could be realized
for realistic values of the current quark masses remained, however,
unanswered. The reason was the missing link between the universal
behavior near $T_c$ and zero current quark mass on one hand and the
known physical properties at $T=0$ for realistic quark masses on the
other hand. Lattice QCD seems particularly suitable for such a 
study, however, exploring the universal region 
is limited by present computer resources \cite{Ukawa}. 

It is the purpose of this talk to provide \cite{BJW} the 
``missing link''.  
Our approach is based on the use of a non--perturbative 
flow equation for a scale dependent effective action \cite{Wet93-2} 
$\Gamma_k$, which is the generating 
functional of the $1 PI$ Green
functions in the presence of an infrared cutoff $\sim k$. 
Varying the infrared cutoff $k$ allows us to consider the relevant
physics in dependence on some momentum--like scale. The standard
effective action is 
obtained by removing the infrared cutoff ($k\ra0$) in the end.
The $k$--dependence of the effective average
action is given by an exact flow
equation~\cite{Wet93-2}, which for scalar fields $\Phi_i$ reads
\begin{equation}
 k \frac{\prl}{\prl k} \Gm_k [\Phi] =
 \hal\Tr\left\{\left(
 \Gm_k^{(2)}[\Phi]+R_k\right)^{-1}
 k \frac{\prl R_k}{\prl k}\right\}
 \label{ERGE}.
\end{equation}
Here $\Gamma_k^{(2)}$ denotes the matrix of second functional
derivatives of $\Gamma_k$ with respect to the field components.
We use a momentum dependent infrared cutoff
$R_k(q)=Z_{\Phi,k}q^2 e^{-q^2/k^2}/(1- e^{-q^2/k^2})$
with $Z_{\Phi,k}$ an appropriate wave function renormalization
constant. In momentum space the trace reads 
${\Tr=\int\frac{d^d q}{(2\pi)^d}\sum_i}$.

We employ for scales below a ``compositeness scale'' of 
$k_{\Phi} \simeq 600 \MeV$ a description in terms of quark and 
scalar mesonic degrees of freedom. This effective quark meson model 
can be obtained from QCD in principle by ``integrating out'' the gluon
degrees of freedom and by introducing fields for composite operators 
\cite{EW94-1}.
In this picture the scale $k_{\Phi}$ is associated
to the scale at which the formation of mesonic bound states can be
observed \cite{EW94-1} in the flow of the momentum dependent four--quark 
interaction. 
We imagine that all other degrees of freedom besides the 
quarks $\psi$ and the scalar and pseudoscalar mesons contained in the 
complex field $\Phi$ are integrated out. Our truncation corresponds
to the ansatz for the effective average action \cite{BJW} 
\begin{equation}
  \label{AAA60}
  \begin{array}{rcl}
  \ds{{\Gamma}_k} &=& \ds{
      \int d^4x\Bigg\{ 
      Z_{\psi,k}\ol{\psi}_a i\slash{\prl}\psi^a+
      Z_{\Phi,k}\tr\left[\prl_\mu\Phi^\dagger\prl^\mu\Phi\right]+
      U_k(\Phi,\Phi^\dagger)
      }\nnn
  \lefteqn{   + \ds{
      \ol{h}_{k}\ol{\psi}^a\left(\frac{1+\gm_5}{2}\Phi_{ab}-
      \frac{1-\gm_5}{2}(\Phi^\dagger)_{ab}\right)\psi^b
      -\frac{1}{2} \tr
      \left(\Phi^\dagger\jmath+\jmath^\dagger\Phi\right)
      \Bigg\} } } \;\;\;\;\; 
  \end{array}
\end{equation}
Here $\Gamma_k$ is invariant under the chiral flavor symmetry
$SU_L(2)\times SU_R(2)$ in absence of the explicit symmetry breaking 
through the source term $\jmath\sim\hat{m}=(m_u+m_d)/2$. 
We will consider the flow
of the most general form of the potential term $U_k$ consistent 
with the symmetries. For $k \to 0$ the potential $U_k$ encodes the 
equation of state. At non--zero temperature
the ansatz \cite{BJW} therefore allows to study the complete 
non--analytic behavior
of the effective potential, or equivalently the free energy, near the 
critical temperature of the second order phase transition.
On the other hand, our 
approximations for the kinetic terms are rather crude and
parameterized by only two running
wave function renormalization constants, $Z_{\Phi,k}$ and
$Z_{\psi,k}$. The same holds for the effective Yukawa coupling
$\ol{h}_k$. We further neglect the scalar
triplet $a_0$ and the pseudoscalar singlet (associated with the
$\eta^\prime$) for $k<k_\Phi$ \footnote{The present
investigation for the two flavor case does not take into account a
speculative ``effective restoration'' of the axial $U_A(1)$ symmetry
at high temperature \cite{PW84-1}.}. This can be achieved in a
chirally invariant way and leads to the $O(4)$ symmetric
linear sigma model for the pions and the sigma resonance, 
however, coupled to quarks now. The non--perturbative flow equations 
for the quark--meson model are obtained from eq.\ (\ref{ERGE})
generalized \cite{Wet90-1} to include fermions and using the 
ansatz (\ref{AAA60}) for 
$\Gamma_k$. We solve them numerically. 

A reliable
quantitative derivation of the effective quark meson model from 
QCD is still missing. We emphasize, however, that the quantitative 
aspects of this derivation will be of minor relevance for our practical 
calculations in the mesonic sector: If the effective Yukawa coupling 
between the quarks and the mesons turns out to be strong at the 
compositeness scale we observe a fast 
approach of the scale dependent effective couplings to approximate
partial infrared fixed points \cite{Ju95-7,BJW}.
As a consequence, the detailed form of the meson potential at 
$k_{\Phi}$ becomes unimportant, except for the value of one relevant 
scalar mass parameter $\overline{m}_{k_{\Phi}}$. Here we fix
$\overline{m}_{k_{\Phi}}$ from phenomenological input such that
$f_{\pi}=92.4 \MeV$ (for $m_{\pi}=135 \MeV$) which sets our unit of
mass for two flavor QCD. The only other input parameter we use is the 
constituent quark mass $M_q$ to determine the scale $k_{\phi}$. 
We consider a range $300 \MeV \lta M_q \lta 350 \MeV$ and find a
rather weak dependence of our results on the precise value of $M_q$.
We point out that though a strong Yukawa coupling at $k_{\Phi}$
is phenomenologically suggested by the comparably large value of 
the constituent quark mass $M_q$ it enters our description as a 
(consistent) assumption.

The equation of state expresses
$\VEV{\ol{\psi}\psi}$ as a function of $T$ and $\hat{m}$ 
where the chiral 
condensate is related to the expectation value $\VEV{\Phi}$ by 
\cite{EW94-1,BJW}
\begin{equation}
\ds{\VEV{\ol{\psi}\psi}} = \ds{
      -2\ol{m}^2_{k_\Phi}\left[
      \VEV{\Phi}-\hat{m}\right]}.
\end{equation}
\begin{figure}
\unitlength1.0cm

\begin{center}
\begin{picture}(8.,3.5)

\put(0.0,0.0){
\epsfysize=8.cm
\rotate[r]{\epsffile{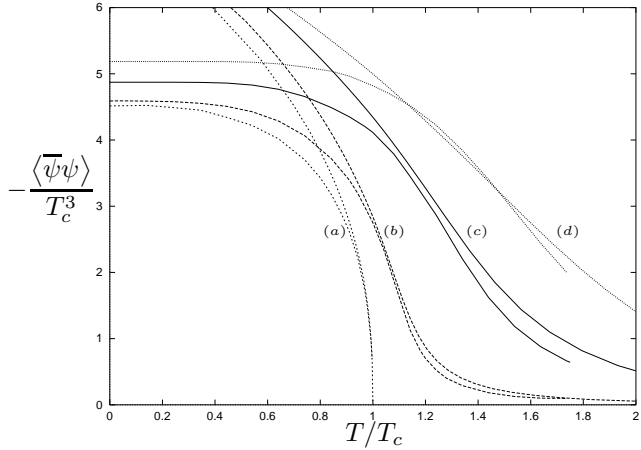}}
}
\put(-0.5,3.){\bf $\ds{-\frac{\VEV{\ol{\psi}\psi}}{T_{c}^3}}$}
\put(4.,-0.25){\bf $\ds{T/T_{c}}$}
\put(3.7,2.5){\tiny $(a)$}
\put(4.5,2.5){\tiny $(b)$}
\put(5.6,2.5){\tiny $(c)$}
\put(6.8,2.5){\tiny $(d)$}
\end{picture}
\end{center}
\caption{\footnotesize The plot shows the chiral condensate
  $\VEV{\ol{\psi}\psi}$ as a function of temperature $T$.  Lines
  $(a)$, $(b)$, $(c)$, $(d)$ correspond at zero temperature to
  $m_\pi=0,45\MeV,135\MeV,230\MeV$, respectively. For each pair of
  curves the lower one represents the full $T$--dependence of
  $\VEV{\ol{\psi}\psi}$ whereas the upper one shows for comparison the
  universal scaling form of the equation of state for the $O(4)$
  Heisenberg model. The critical temperature for zero quark mass is
  $T_c=100.7\MeV$. The chiral condensate is normalized at a scale
  $k_{\Phi}\simeq 620\MeV$.}
\label{ccc_T}
\end{figure}
Curve $(a)$ of figure \ref{ccc_T} gives the temperature 
dependence of $\VEV{\ol{\psi}\psi}$
in the chiral limit $\hat{m}~=~0$. Here the lower curve is the full
result for arbitrary $T$ whereas the upper curve \cite{BJW,BTW95}
corresponds to the
universal scaling form of the equation of state for the $O(4)$
Heisenberg model.  
We see perfect agreement of both curves for $T$
sufficiently close to $T_c=100.7 \MeV$. This demonstrates the
capability of our method to cover the critical behavior and, in
particular, to reproduce~\cite{BJW}
the critical exponents of the $O(4)$--model.
The curves $(b)$, $(c)$
and $(d)$ are for non--vanishing values of the average current quark
mass $\hat{m}$.  Curve $(c)$ corresponds to $\hat{m}_{\rm phys}$ or,
equivalently, $m_\pi(T=0)=135\MeV$. One observes a crossover in the
range $T=(1.2-1.5)T_c$. In order to
facilitate comparison with lattice simulations which are typically
performed for larger values of $m_\pi$ we also present results for
$m_\pi(T=0)=230\MeV$ in curve $(d)$. One may define a ``pseudocritical
temperature'' $T_{pc}$ associated to the smooth crossover as the
inflection point of $\VEV{\ol{\psi}\psi}(T)$. 
The value for the pseudocritical temperature for $m_{\pi}=230 \MeV$
is $T_{pc}\simeq 150 \MeV$. For realistic quark mass, or 
$m_\pi = 135 \MeV$, we obtain $T_{pc}\simeq 130 \MeV$. 

\begin{figure}
\unitlength1.0cm
\begin{center}
\begin{picture}(8.,3.5)

\put(0.0,0.0){
\epsfysize=8.cm
\rotate[r]{\epsffile{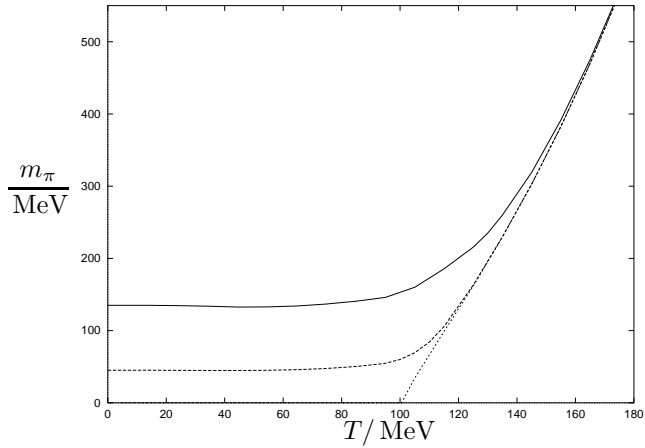}}
}
\put(-0.5,3.){\bf $\ds{\frac{m_\pi}{\MeV}}$}
\put(4.,-0.25){\bf $\ds{T/\MeV}$}

\end{picture}
\end{center}
\caption{\footnotesize The plot shows $m_\pi$ as a function of
  temperature $T$ for three different values of the average light
  current quark mass $\hat{m}$. The solid line corresponds to the
  realistic value $\hat{m}=\hat{m}_{\rm phys}$ whereas the dotted line
  represents the situation without explicit chiral symmetry breaking,
  i.e., $\hat{m}=0$. The intermediate, dashed line assumes 
  $\hat{m}=\hat{m}_{\rm phys}/10$.}
\label{mpi_T}
\end{figure}

A second important result of our investigations is the temperature
dependence of the space--like pion correlation length
$m_\pi^{-1}(T)$. The plot for $m_\pi(T)$ in figure \ref{mpi_T} again 
shows the second order phase transition in the chiral limit $\hat{m}=0$. 
In this limit the behavior for small positive $T-T_c$ is 
characterized by the critical exponent $\nu$, i.e.
$m_\pi(T)=\left(\xi^+\right)^{-1}T_c \left( (T-T_c)/T_c\right)^\nu$
and we obtain $\nu=0.787$, $\xi^+=0.270$. For $\hat{m}>0$ we find that
$m_\pi(T)$ remains almost constant for $T\lta T_c$. 
For $T>T_c$ the correlation length
decreases rapidly and for $T\gg T_c$ the precise value of $\hat{m}$
becomes irrelevant. The overall size
of the pion correlation length near the critical temperature 
is given by $ m_\pi(T_{pc})\simeq 1.7 m_\pi(0)$ for the realistic value
$\hat{m}_{\rm phys}$. 

We point out two important answers one obtains from this 
study:
First of all, for a thermal equilibrium situation the chiral  
transition gives no indication for strong fluctuations of
pions with long wavelength 
\footnote{See ref.\ \cite{RaWi93-1} for a 
discussion in the context 
of a chiral transition far from equilibrium.}. The longest correlation
lengths near the crossover temperature $T_{pc}$ even becomes 
smaller than at ${T=0}$~\footnote{It should be emphasized, however, that  
a tricritical behavior with a massless excitation remains possible for
three flavors \cite{GGP2}.}. The second answer concerns the 
applicability of universal, i.e.\ almost model independent, arguments 
for a description of the chiral phase transition.
Despite the observed comparably short 
correlation lengths at non--zero temperature, we find the approximate
validity of the $O(4)$ scaling behavior over a large temperature interval
near and above $T_c$ even for quark masses somewhat
larger than the realistic ones. The present approach can be extended
to the case with three light quark flavors \cite{BJW97-2}.

\section*{Acknowledgments}
I want to thank D.-U.\ Jungnickel and C.\ Wetterich for 
collaboration on the subject presented in this talk. 
I am grateful to the organizers of this conference for
providing a most stimulating environment.


\vspace*{-0.2cm}
\section*{References}
\begin{thebibliography}{10}
\bib{CP75} J.C.\ Collins and M.J.\ Perry, Phys.\ Rev.\ Lett.\ {\bf 34}
(1975) 1353.
\bibitem{PW84-1} R.D. Pisarski and F. Wilczek, \PR{D29} (1984) 338.
\bibitem{RaWi93-1} K. Rajagopal and F. Wilczek, \NP{B399} (1993) 395;
  {\bf B404} (1993) 577; K.~Rajagopal, in {\em Quark-Gluon Plasma 2}, 
  edited by R.\ Hwa (World Scientific, 1995) (hep-ph/9504310).
\bibitem{Ans88-1} See e.g.\ J.~D.~Bjorken, K.~L.~Kowalski and
  C.~C.~Taylor, preprint SLAC-PUB-6109 (hep-ph/9309235) and references
therein.
\bib{Ukawa} For a review see A.\ Ukawa, these proceedings.
\bib{BJW} J.\ Berges, D.-U.\ Jungnickel and C.\ Wetterich,
preprint HD--THEP--97--20 (hep-ph/9705474).
\bib{Wet93-2} C.\ Wetterich, \plb{301} (1993) 90 and references
therein.
\bibitem{EW94-1} U. Ellwanger and C. Wetterich, Nucl.\ Phys.\
{\bf B423} (1994) 137.
\bibitem{Wet90-1} C.~Wetterich, \ZP{C48} (1990) 693.
\bib{Ju95-7} D.-U.\ Jungnickel and C.\ Wetterich, \prd{53} (1996)
5142.
\bib{BTW95} J.\ Berges, N.\ Tetradis and C.\ Wetterich, 
Phys.\ Rev.\ Lett.\ {\bf 77} (1996) 873.
\bib{GGP2} S.\ Gavin, A.\ Gocksch and R.D.\ Pisarski,
Phys.\ Rev.\ {\bf D49} (1994) 3079. 
\bib{BJW97-2} J.\ Berges, D.-U.\ Jungnickel and C.\ Wetterich, 
in preparation.
\end{thebibliography}
\end{document}